\documentclass[prd, twocolumn, aastex, superscriptaddress]{revtex4-1}
\pdfoutput=1 
\usepackage{amsmath,amstext}
\usepackage[T1]{fontenc}
\usepackage{apjfonts} 
\usepackage{graphicx}
\usepackage{natbib}
\usepackage{xcolor}
\usepackage{hyperref}

\usepackage{aasjournal}

\begin{document}

\title{CMB distance priors revisited: effects of dark energy dynamics, spatial curvature, \\
primordial power spectrum, and neutrino parameters}

\author{Zhongxu Zhai}
\email{zhai@ipac.caltech.edu}
\affiliation{IPAC, California Institute of Technology, Mail Code 314-6, 1200 E. California Blvd., Pasadena, CA 91125, USA}

\author{Chan-Gyung Park}
\affiliation{Division of Science Education and Institute of Fusion Science, Jeonbuk National University, Jeonju 54896, Republic of Korea}

\author{Yun Wang}
\affiliation{IPAC, California Institute of Technology, Mail Code 314-6, 1200 E. California Blvd., Pasadena, CA 91125, USA}

\author{Bharat Ratra}
\affiliation{Department of Physics, Kansas State University,
116 Cardwell Hall, Manhattan, KS 66506, USA}

\begin{abstract}
As a physical and sufficient compression of the full CMB data, the CMB distance priors, or shift parameters, have been widely used and provide a convenient way to include CMB data when obtaining cosmological constraints. In this paper, we revisit this data vector and examine its stability under different cosmological models. We find that the CMB distance priors are an accurate substitute for the full CMB data when probing dark energy dynamics. This is true when the primordial power spectrum model is directly generalized from the power spectrum of the model used in the derivation of the distance priors from the CMB data. We discover a difference when a non-flat model with the untilted primordial inflation power spectrum is used to measure the distance priors. This power spectrum is a radical change from the more conventional tilted primordial power spectrum and violates fundamental assumptions for the reliability of the CMB shift parameters. We also investigate the performance of CMB distance priors when the sum of neutrino masses $\sum m_{\nu}$ and the effective number of relativistic species $N_{\text{eff}}$ are allowed to vary. Our findings are consistent with earlier results: the neutrino parameters can change the measurement of the sound horizon from CMB data, and thus the CMB distance priors. We find that when the neutrino model is allowed to vary, the cold dark matter density $\omega_{c}$ and $N_{\text{eff}}$ need to be included in the set of parameters that summarize CMB data, in order to reproduce the constraints from the full CMB data. We present an updated and expanded set of CMB distance priors which can reproduce constraints from the full CMB data within $1\sigma$, and are applicable to models with massive neutrinos, as well as non-standard cosmologies.
\end{abstract}


\maketitle

\section{Introduction}

The cosmic microwave background (CMB) radiation contains important information about the evolution of the universe, as well as about the formation of the structures we observe today. Accurate measurements made in the past two decades, through experiments such as WMAP \cite{Spergel_2003} and Planck \cite{Planck_2013}, provide interesting constraints on the matter component in the universe, on spatial curvature, on dark energy properties, on neutrino masses, on inflation models, and so on. The analysis is usually done by comparing the observed power spectrum of the temperature fluctuations with predictions from a particular cosmological model. This requires the full knowledge of the linear perturbation theory predictions of the model, either analytical or numerical. For models like $\Lambda$CDM or minimal extensions, this approach has been demonstrated to be quite successful. And the combination of CMB observations with other cosmic measurements such as type Ia supernova apparent magnitudes \cite{Scolnic_2018}, baryon acoustic oscillation peak scales \cite{Alam_2017}, and Hubble parameter data \cite{Farooq_2017}, can further improve the cosmological parameter constraints and help to break degeneracies between cosmological parameters. However, for some models with exotic properties, especially cases where dynamical dark energy is introduced to explain the observed current cosmic acceleration, the complete modeling of the linear perturbation results can be non-trivial and challenging. This problem can be more severe for models where the underlying gravity theory is also modified \cite{Ishak_2019}.

One possible solution from the observational side is to express the CMB data in a more concise and model-independent way. In this paper, we focus on the so-called CMB distance priors which can represent the full CMB data in terms of only a few data points. This method has been widely applied and provides useful constraints on the cosmological model. The standard CMB distance priors are composed of two CMB shift parameters $R$ and $l_{a}$. The first shift parameter $R$ is observed to be independent of the underlying models if they have the same baryonic and dark matter components and primordial fluctuation spectrum \cite{Efstathiou_1999}. The second shift parameter $l_{a}$, proposed in \cite{Wang_2006, Wang_2007}, is nearly uncorrelated with $R$ and therefore these two parameters can be used as a numerically economical compression of the full CMB data. CMB distance priors can provide constraints on some of dark energy models that are consistent with the constraints derived using the full CMB data. They also can provide consistent constraints on more exotic models for both dark energy and modified gravity. We refer the reader to Refs. \cite{Lazkoz_2006, Nesseris_2007, Fairbairn_2007, Sollerman_2009, Wu_2010, WangS_2013, Wang_2013, Aubourg_2015, Zhai_2017, Chen_2018, Zhai_2019, Arjona_2019, Rezaei_2019} and references therein for examples.

Although the CMB distance priors or the shift parameters have been used to produce important and informative cosmological results, we should be careful when using this compressed data set and must determine how accurately it represents the full CMB data. By definition, the CMB shift parameters are perhaps the least model-dependent parameters that can be extracted from the CMB data \cite{Wang_2006}. However, their determination is based on the assumption of a specific model, such as the flat $\Lambda$CDM model or a simple extension of this model. Therefore the CMB shift parameters are not directly measured quantities and this can invalidate their usage in constraining an arbitrary model. If not applied properly, the CMB shift parameters may give biased result under certain conditions. This issue was studied in Ref.\ \cite{Elgaroy_2007}, where the authors compared values of CMB shift parameters determined using various cosmological models and identified some parameter ranges over which the values of the shift parameters were relatively stable. Similar work was conducted in Ref.\ \cite{Corasaniti_2008}, where the authors found that the CMB shift parameters are stable under minimal modifications of the dark energy parameters, but some percent change is found when the models are generalized to allow the sum of neutrino masses or the primordial power spectrum to vary. These analyses investigated the robustness of the CMB shift parameters and clarified some sources of possible confusion and bias on the cosmological parameter constraints that may result from the introduction of such additional cosmological variables. We note that other methods for compressing CMB data have been proposed. For instance, the authors of Ref.\ \cite{Prince_2019} apply a Massively Optimized Parameter Estimation technique to compress the full CMB data into a vector which has the same dimension as the model parameters of interest, which can also reduce the computational demand significantly when incorporating CMB data in cosmological model analyses.

Significant improvements in data quality and quantity have resulted in 
studies of larger model parameter spaces, with a larger number of cosmological parameters now in the mix. It is necessary and timely to explicitly and quantitatively reexamine the performance of CMB data compression. In this paper, we focus on and revisit the robustness of CMB shift parameters. In particular, we use cosmological models with different physical parameters (dark energy dynamics parameters, spatial curvature, primordial power spectrum parameters, as well as the sum of neutrino masses and the number of relativistic species) to measure the CMB shift parameters and study their robustness. Based on our results we also propose a method to generalize and improve the CMB distance priors so that the bias in the resulting cosmological parameter constraints are minimized.

This paper is organized as follows. In Sec.\ \ref{sec:CMBpriors}, we describe the basics of CMB distance priors. In Sec.\ \ref{sec:model}, we introduce the cosmological models considered in our analysis. Our results are presented in Sec.\ \ref{sec:results}. Section \ref{sec:conclusion} contains our discussion and conclusion.

\section{CMB distance priors} \label{sec:CMBpriors}

The spacetime model we consider in this paper is based on a FLRW metric, under which the comoving distance to an object at redshift $z$ is given by
\begin{eqnarray}\label{eq:comodis}
    r(z) = \frac{c}{H_{0}}|\Omega_{k}|^{-1/2}\text{sinn}[|\Omega_{k}|^{1/2}\Gamma(z)], \\
    \Gamma(z) = \int_{0}^{z}\frac{dz'}{E(z')}, \quad E(z) = H(z)/H_{0},
\end{eqnarray}
where $c$ is the speed of light, $H_{0}=100 h\text{ km s}^{-1}\text{Mpc}^{-1}$ with $h$ the dimensionless Hubble constant, $H(z)$ is the Hubble parameter, and $\text{sinn}(x)=\sin(x), x, \text{and } \sinh(x)$ for the current value of the spatial curvature density parameter $\Omega_{k}<0, \Omega_{k}=0, \text{and } \Omega_{k}>0$, respectively. The CMB shift parameters are defined as 
\begin{eqnarray}
    R &\equiv&\sqrt{\Omega_{m}H_{0}^{2}} \, r(z_{*})/c, \\
    l_{a} &\equiv& \pi r(z_{*})/r_{s}(z_{*}),
\end{eqnarray}
where $z_{*}$ is the redshift to the photon-decoupling surface, $r_s$ is the sound horizon, and $\Omega_{m}$ is the current value of the fraction of non-relativistic matter in the universe. These two CMB shift parameters together with the current value of the physical fraction of baryonic matter $\omega_{b}=\Omega_{b}h^{2}$ and spectral index of the primordial power spectrum $n_{s}$ can give an efficient summary of CMB data for cosmological parameters constraint purposes. These quantities are also known as CMB distance priors. We focus on the geometrical probe of CMB data in this work, so we ignore $n_{s}$ in the following analysis and only present results for the vector formed by $(R, l_{a}, \omega_{b})$. In addition, we determine the CMB distance priors in this work by using CAMB \cite{Lewis_2002ah} due to the complexity of the models. The fitting formula used to calculate the CMB related parameters such as $r(z_{*})$ and the sound horizon may have percent-level differences from the correct values \cite{Mehta_2012}. However, for the purpose of testing the CMB distance priors, the results are not expected to change significantly as long as the computation is self-consistent.

\section{cosmological models}\label{sec:model}

In this section, we introduce the models we use to study how various cosmological parameters impact the usage of CMB distance priors. These parameters include those that govern the dark energy dynamics, spatial curvature and different primordial power spectra, as well as the neutrino parameters that can be constrained by the CMB data. We summarize the models considered in this work in Table \ref{Table:Model}.

\begin{table}\label{tab:models}
\centering
\begin{tabular}{llll}
\hline
Model   & Acronym & parameter\footnote{For models other than flat $\Lambda$CDM, only the additional parameters are listed. $\Omega_c$ is the current value of the cold dark matter density parameter. Other symbols are defined in the main text.}   \\
\hline
Flat $\Lambda$CDM model& $\Lambda$CDM  & $\Omega_{b}$, H$_{0}$, $\Omega_{c}$ \\
Non-flat $\Lambda$CDM model & $o\Lambda$CDM  & $\Omega_{k}$   \\
Flat $w$CDM model & $w$CDM & $w$ \\
Non-flat $w$CDM model & $ow$CDM & $w$, $\Omega_{k}$ \\
Flat $\phi$CDM model & $\phi$CDM & $\alpha$ \\
Non-flat $\phi$CDM model & $o\phi$CDM & $\alpha$, $\Omega_{k}$ \\
running scalar index with tensor mode & $\Lambda$CDM+$n_{\rm run}$+$r$ & $n_{\rm run}$, $r$ \\
log oscillation & $\Lambda$CDM+LO & $A_{\text{log}}$, $\omega_{\text{log}}$, $\phi_{\text{log}}$\\
Neutrino model & $\nu$CDM & $\sum m_{\nu}(\text{eV})$, $N_{\text{eff}}$\\
\hline
\end{tabular}
\caption{Models considered in the paper, including their main characters, names and parameters. All models except $\nu$CDM adopt  $\sum m_{\nu}$ = 0.06 eV and the standard neutrino content $N_{\text{eff}} = 3.046$.}
\label{Table:Model}
\end{table}

\subsection{Dark energy models}

 The Hubble parameter $H(z)$ is given by
\begin{equation}
    H(z)^2 = H_{0}^{2}[\Omega_{m}(1+z)^3+\Omega_{r}(1+z)^4 + \Omega_{k}(1+z)^2 + \Omega_{X}X(z)],
\end{equation}
with constraint $\Omega_{m}+\Omega_{r}+\Omega_{k}+\Omega_{X}=1$. The function $X(z)\equiv\rho_{X}(z)/\rho_{X}(0)$ describes the time evolution of the dark energy density. The current value of the radiation density parameter $\Omega_{r}=\Omega_{m}/(1+z_{\text{eq}})\ll\Omega_{m}$ and can be omitted in late time cosmology studies (here $z_{\text{eq}}$ is the redshift of matter-radiation equality).  

The cosmological constant corresponds to a constant $\rho_{X}$. For comparison with this simple model of dark energy, we also consider parameterizations of the dark energy equation of state parameter $w(z)$
\begin{equation}
    \frac{\rho_{X}(z)}{\rho_{X}(0)}=\exp{\left(3\int_{0}^{z}\frac{1+w(z')}{1+z'}dz'\right)}.
\end{equation}
In particular we assume a constant $w(z)$ which corresponds to the $w$CDM or XCDM model \cite{Turner_1997, Chiba_1997}. This is not a physically consistent dynamical dark energy model.

We also consider a scalar field dark energy model which is the simplest physically consistent dynamical dark energy model \cite{Peebles_1988, Ratra_1988, Pavlov_2013}. In this model, the dynamical properties of dark energy are determined by the potential energy density of the scalar field:
\begin{equation}
    V(\phi)=V_{0}\phi^{-\alpha},
\end{equation}
where $V_{0}$ is the amplitude of the potential and $\alpha$ is a constant parameter to be determined by observations. The evolution of the scalar field is obtained by solving the Klein-Gordon equation
\begin{equation}
   \phi'' + \left(3+\frac{\dot{H}}{H^2}\right)\phi' {-\hat{V}_{0}\alpha\phi^{-\alpha-1} } \left(\frac{H_{0}}{H}\right)^2=0,
\end{equation}
where $\phi'\equiv d\phi/d \ln{a}$, $H=\dot{a}/a$, $\hat{V}_0 \equiv V_0/H_{0}^2$, and an overdot denotes the time derivative $d/dt$. The expansion of the universe in this model can be computed from 
\begin{eqnarray}
    & & \left(\frac{H}{H_{0}}\right)^2 = \\ 
    & & \frac{1}{1-\frac{1}{6}(\phi')^2}\left[\Omega_{m}(1+z)^3  {+\Omega_r (1+z)^4} + \Omega_{k}(1+z)^2+ {\frac{1}{3}\hat{V}_{0}\phi^{-\alpha} }  \right], \nonumber
\end{eqnarray}
where we have chosen units such that the Newtonian gravitational constant $G =1/(8\pi)$.

\subsection{Primordial power spectrum and spatial curvature}\label{sec:power}

In a spatially flat universe, non-slow-roll (tilted) inflation generates a primordial scalar energy density inhomogeneity power spectrum \cite{Lucchin_1985, Ratra_1992, Ratra_1989} 
\begin{equation}\label{eq:tilted}
    P_{0}(k) = A_{s}\left(\frac{k}{k_{0}}\right)^{n_{s}},
\end{equation}
where $k$ is the wave number, $A_{s}$ is the amplitude at a pivot scale {$k_{0}=0.05~\text{Mpc}^{-1}$}, and $n_{s}$ is the power spectral index. 

On the other hand, in a non-flat universe, the spatial curvature introduces an additional length scale and likely invalidates the above tilted primordial power spectrum. In the slow-roll (untilted) non-flat inflation model, the primordial power spectrum \cite{Ratra_1995, Ratra_2017} is 
\begin{equation}\label{eq:untilted}
    P(q)\propto\frac{(q^2-4K)^2}{q(q^2-K)}
\end{equation}
where the wave number $q=\sqrt{k^2+K}$ and $K=-(H_{0}^2/c^2)\Omega_{k}$ is the spatial curvature. This power spectrum can be normalized at the pivot scale $k_{0}$ with the amplitude $A_{s}$ in a similar manner as in the spatially flat case. Note that this model is not a simple generalization of the flat case. When $\Omega_{k}=0$, it is equivalent to $n_{s}=1$ in the flat universe. Therefore these two models are not ``nested": one model is not a special class of the other. CMB data \cite{Ooba_2018a, Ooba_2018b, Ooba_2018c, Park_2019a, Park_2019b, Park_2018, Handley_2019, Park_2019c} and other cosmological observations \cite{Farooq_2015, Yu_2016, Li_2016, Wei_2017, Rana_2017, Yu_2018, Park_2019d, Abbott_2019, Mitra_2019, Ryan_2019, Li_2019, Jesus_2019, Khadka_2019} are not inconsistent with a mildly closed cosmological model. 

In addition, we also consider some extended models based on the tilted model power spectrum of Eq.\ (\ref{eq:tilted}). The first model has a running scalar spectral index
\begin{equation}
{P(k) = A_{s}\left(\frac{k}{k_{0}}\right)^{n_{s}+\frac{1}{2}n_{\rm run} \ln(k/k_{0})}, }
\end{equation}
where $n_{\rm run} = \text{d}n_{s}/\text{d}\ln{k}$. The running of the running of the scalar spectral index is straightforward to implement in this model. In addition, we also consider the effect of primordial gravitational waves, or the tensor mode, which can be parameterized by the tensor-to-scalar ratio $r$. We adopt the same method as the Planck team \cite{Planck_2018_6} in this analysis, which assumes a pivot scale at 0.002 Mpc$^{-1}$.
Current CMB observational data also allows examination of more radical departures from the fiducial power-law model. As an example, we adopt a parameterized model of a logarithmic oscillation \cite{Martin_2001, Danielsson_2002, Bozza_2003}
\begin{equation}
    P^{\text{log}} = P_{0}(k)\left\{1+A_{\log}\cos\left[ \omega_{\log} { \ln{\left(\frac{k}{k_{0}}\right)} }+\phi_{\log}\right] \right\},
\end{equation}
where $A_{\log}, \omega_{\log}$ and $\phi_{\log}$ are model parameters. We note that this model was also investigated by the Planck team \cite{Planck_2016_inflation}.

\subsection{Neutrino parameters}

In addition to the tests with dark energy models, spatial curvature and primordial power spectrum, we consider the effect of neutrino properties which are also an important objective of precision cosmology. The latest CMB measurements tightly constrain the sum of neutrino masses $\sum m_{\nu}<0.54$ $\text{eV}$ \cite{Planck_2018_6} when only the high $\ell$ (multipole number) temperature information is considered. Adding information from lensing and observational data from large scale structure through baryon acoustic oscillation (BAO) data, the constraint is significantly tightened to be $\sum m_{\nu}<0.13~\text{eV}$ at 95\% confidence level in this minimal model considered by Ref. \cite{Planck_2018_6}. In addition, the neutrino mass has a direct correlation with the value of $H_{0}$ \cite{Riess_2018a}. Therefore accurate and unbiased measurement of neutrino mass is of critical importance for multiple reasons \cite{Vagnozzi_2017}. 

The effective number of relativistic species $N_{\text{eff}}$ is defined through the total relativistic energy density well after electron-positron annihilation \cite{Lesgourgues_2014}
\begin{equation}
     \rho_{\text{rad}} = \left[1+N_{\text{eff}}\frac{7}{8}\left(\frac{4}{11}\right)^{4/3}\right]\rho_{\gamma},
\end{equation}
where $\rho$ is the energy density, and the subscript "$\gamma$" and "rad" refers to the contribution from photons and radiation respectively. The standard cosmological model predicts that $N_{\text{eff}}\approx3.046$, however deviation can be easily introduced in exotic cosmological models, for instance through dark radiation \cite{Randall_1999, Gaodlowski_2006}. This parameter has a direct impact when the universe is radiation dominated and thus changes the sound horizon. Similarly as for the neutrino mass, $N_{\text{eff}}$ is also degenerate with $H_{0}$ and a larger value can alleviate the tension between $H_{0}$ measured from Planck and from some local distance ladder observations \cite{Calabrese_2012, Planck_2018_6}. We note that a number of other $H_0$ measurements are more consistent with (but slightly larger than) the Planck estimate \cite{Gott_2001, Chen_2003, Chen_2011, Chen_2017, Lin_2017, Abbott_2018, GomezValent_2018, Haridasu_2018, Zhang_2018, Dominguez_2019, Cuceu_2019, Zeng_2019} and that some local expansion rate estimates are lower with larger error bars and so less inconsistent with the Planck $H_0$ value \cite{Rigault_2015, Zhang_2017, Dhawan_2018, FernandezArenas2018, Freedman_2019}, but see Ref.\ \cite{Yuan_2019}.  

\section{Results}\label{sec:results}

\subsection{Performance of CMB distance priors}

In this analysis, we use the Planck 2015 release data \cite{Planck_2015} for the extraction of CMB distance priors. We note that these data are superseded by the final Planck 2018 release data \cite{Planck_2018_6} and the cosmological constraints we derive here can be improved, but the results we infer about the CMB distance priors are not expected to change significantly. In particular, we use the \textsf{TT+lowP+lensing} CMB data in our computations here, but it is straightforward to use other Planck CMB data combinations. We give the expanded CMB distance priors for both Planck 2015 and Planck final data in Sec.\ref{sec:neutrinos}.

\begin{figure}[htbp]
\begin{center}
\includegraphics[width=9.0cm]{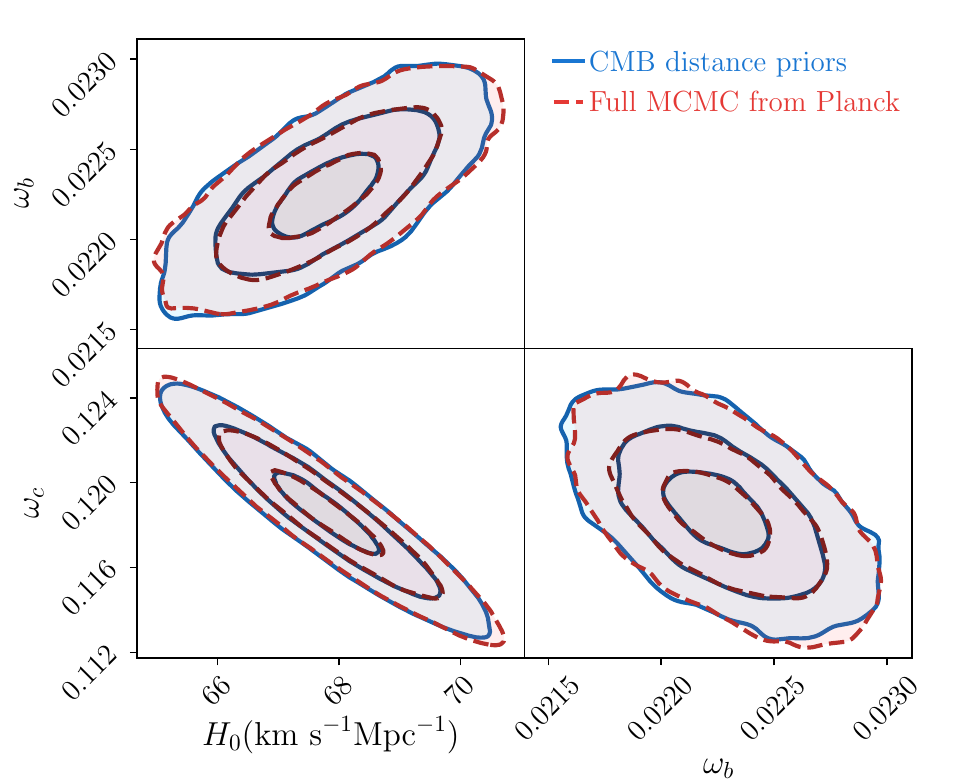}
\caption{1, 2 , and 3$\sigma$ constraints on the parameters of the spatially flat $\Lambda$CDM model from the full CMB data (red dashed contours) and from the CMB distance priors (blue solid contours).} 
\label{fig:LCDM_con}
\end{center}
\end{figure}

Figure \ref{fig:LCDM_con} is an illustrative example of the reproduction of the constraints determined from the full CMB data by those derived using just the CMB distance priors. We assume a spatially flat $\Lambda$CDM to extract the CMB distance priors from the corresponding MCMC analysis results, and then use these  compressed data to constrain the cosmological model.

\begin{figure}[htbp]
\begin{center}
\includegraphics[width=8.5cm]{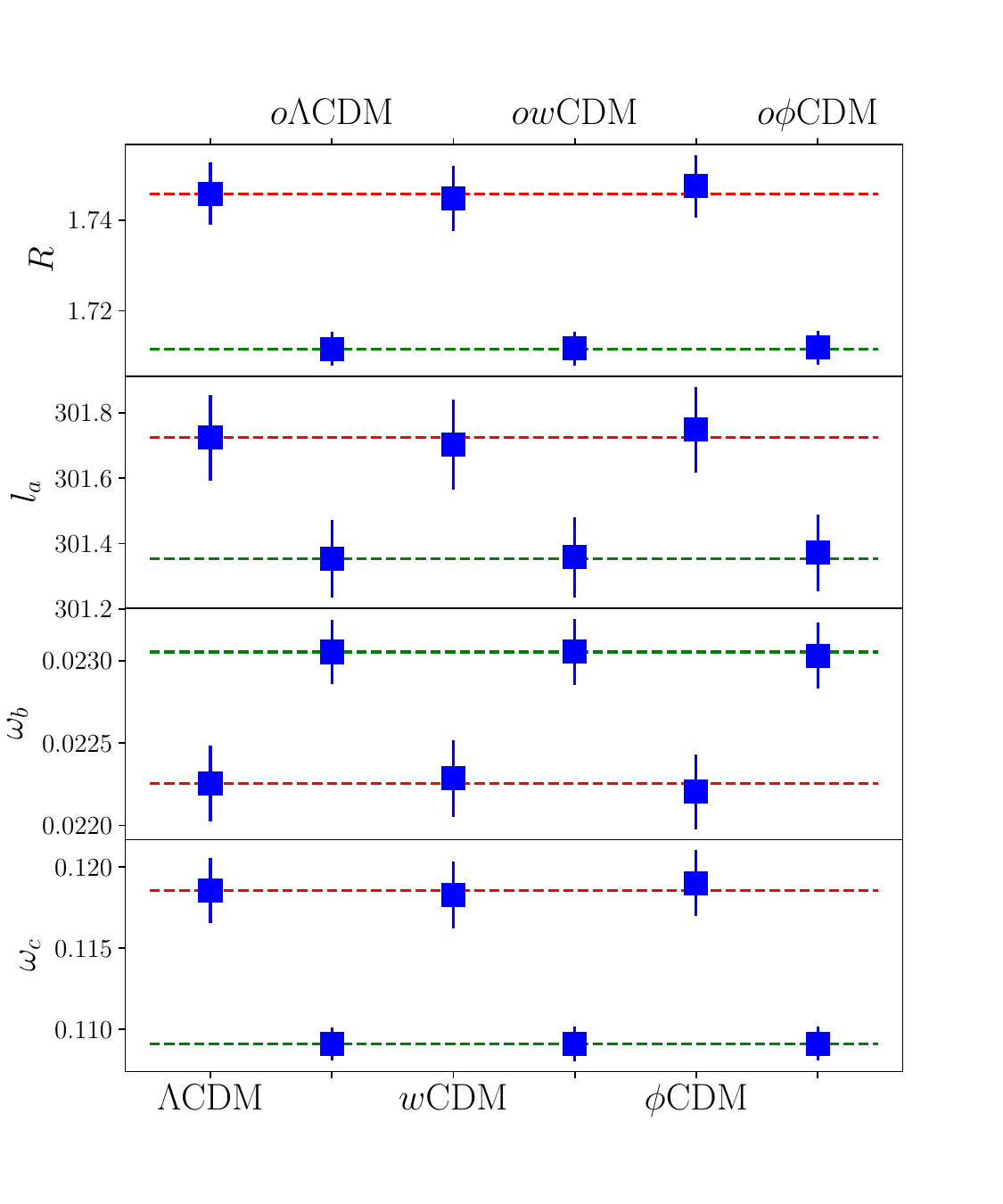}
\includegraphics[width=8.5cm]{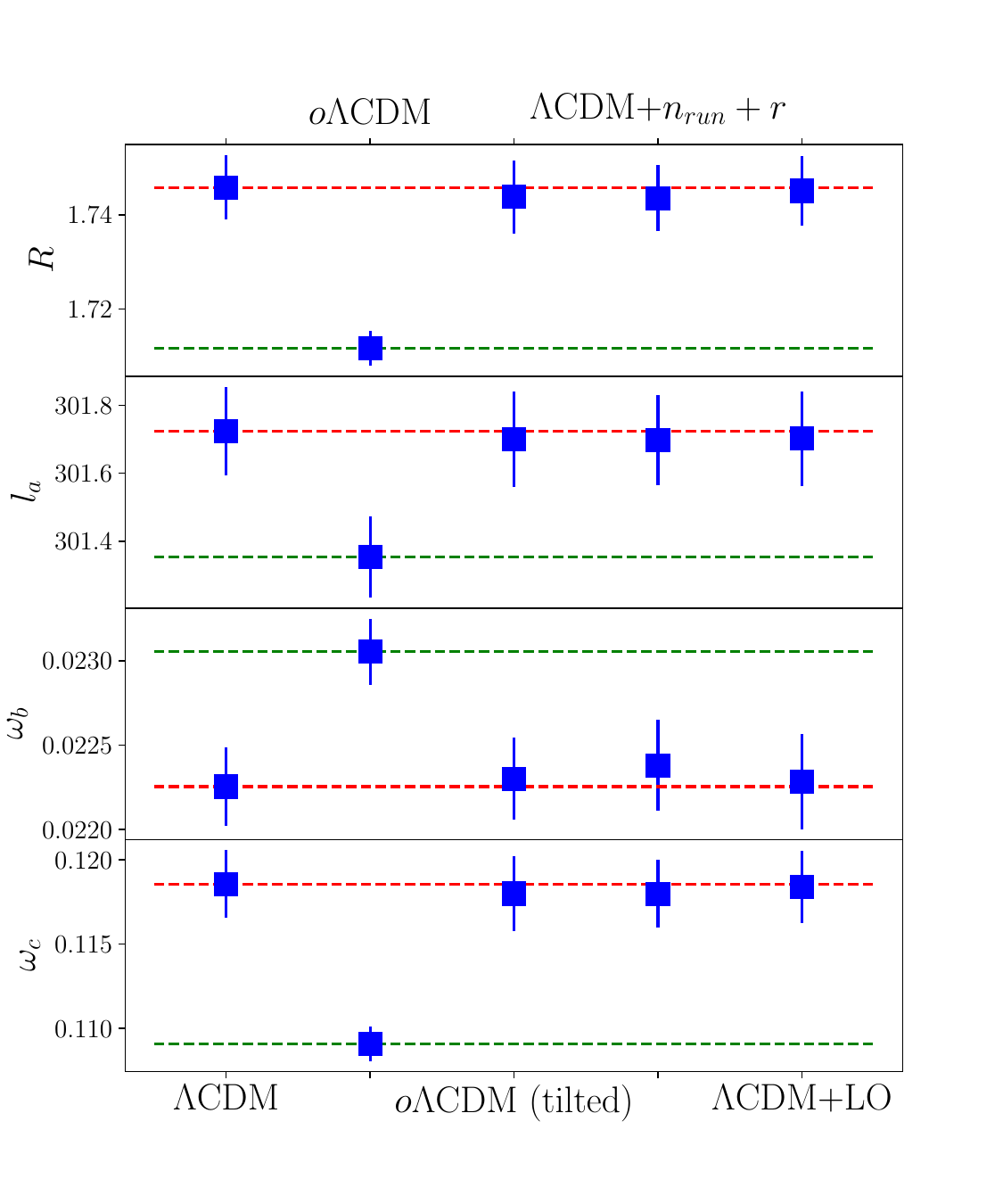}
\caption{The CMB distance priors $R, l_{a}$ and $\omega_{b}$ extracted by assuming different models as indicated on the $x$-axes, but from the same input CMB data. The bottom panel shows the cold dark matter fraction $\omega_{c}$, see Sec.\ \ref{sec:neutrinos} for a discussion of this parameter. The red dashed line is the result determined using the spatially flat $\Lambda$CDM model with tilted primordial power spectrum, while the green dashed line uses the non-flat $\Lambda$CDM model with untilted primordial power spectrum. $Top$ $panel:$ comparison of results from various dark energy models. $Bottom$ $panel:$ comparison of results from various primordial power spectrum models; results from the non-flat $\Lambda$CDM model with tilted primordial power spectrum (used in the Planck group analysis of non-flat models) are also shown.}
\label{fig:CMBshift}
\end{center}
\end{figure}

In order to investigate the stability of CMB shift parameters under various conditions, we use the full CMB data to constrain the models described in Sec.\ \ref{sec:model} by running a full MCMC analysis. We then extract the CMB distance priors based on this analysis for each model and compare the results in Fig.\ \ref{fig:CMBshift}. The top panel is a comparison of the dark energy models, and the effect of excluding or including spatial curvature. Note that the primordial power spectrum for the non-flat models is the untilted one given in Eq.\ (\ref{eq:untilted}). We see that the CMB distance priors are independent of the model of dark energy assumed, consistent with the findings in Ref.\ \cite{Wang_2007}. On the other hand, they do depend significantly on the assumed power spectrum with the flat tilted model results differing from the non-flat untilted model ones.  
For the flat tilted and non-flat untilted inflation models (with power spectra given in Eqs.\ (\ref{eq:tilted}) and (\ref{eq:untilted})), the spatially flat model is not nested within the non-flat model. This causes some parameter values in the non-flat model to deviate from those in the flat model, which alters the CMB shift parameters. This is consistent with the large difference of the $\chi^{2}$ minimum values in the MCMC results of the flat and non-flat models.

The above result reveals the necessity to further investigate the modeling of physics in the early universe. We first focus on the primordial power spectra described in Sec.\ \ref{sec:power}. With the same CMB data, we run a MCMC analysis for various models of primordial power spectrum and the extracted CMB shift parameters are shown in the bottom panel of Fig.\ \ref{fig:CMBshift}. The first two models are the flat $\Lambda$CDM model with tilted primordial power spectrum (Eq.\ref{eq:tilted}) and non-flat $\Lambda$CDM model with untilted primordial power spectrum (Eq.\ref{eq:untilted}) as reference, the same as in the top panel of Fig.\ \ref{fig:CMBshift}. The tests show that CMB shift parameters are stable under direct generalization of the tilted primordial power spectrum, including extensions to a tilted model with non-zero spatial curvature, running of the spectral index and including the tensor mode, and phenomenological parameterizations. The variation for all the observables in the CMB distance priors are significantly lower than $1\sigma$. Combined with the tests on the dark energy models, we can see that for small deviation or generalization from fiducial $\Lambda$CDM model, expressing the full CMB data in terms of the compressed form of CMB distance priors can result in unbiased and consistent constraints. For models with a radical difference from the $\Lambda$CDM model with a tilted primordial power spectrum, the extracted CMB distance priors are systematically offset. This indicates that caution is warranted when using CMB distance priors, instead of the full CMB data in a data analysis, since the resulting parameter constraints could be significantly biased, possibly by as much as a few $\sigma$.

\begin{figure}[htbp]
\begin{center}
\includegraphics[width=9.0cm]{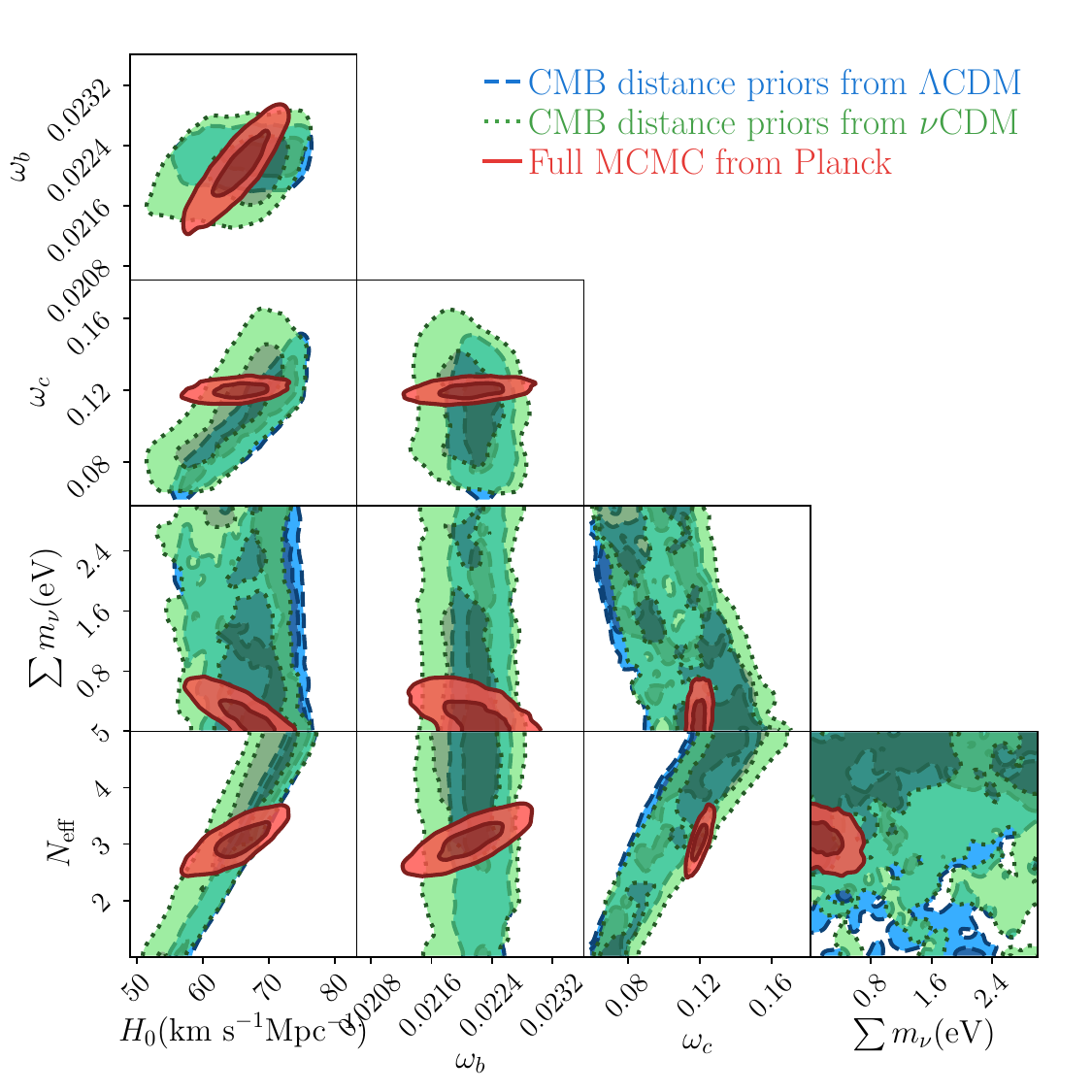}
\caption{Constraints on the $\nu$CDM model parameters derived using the CMB distance priors, compared with those determined from the full CMB data. The blue contours assume CMB distance priors measured using the flat $\Lambda$CDM model, while the green contours assume distance priors extracted using the $\nu$CDM model. The big discrepancy with the result derived using the full CMB data (red contours) shows that the usual CMB distance priors are insufficient for constraining the $\nu$CDM model regardless of how the priors are computed. }
\label{fig:nuCDM_con_old}
\end{center}
\end{figure}

Next we investigate the effect of neutrino parameters on the determination of the CMB distance priors. This was first explored in Ref.\ \cite{Elgaroy_2007} where the authors incorporate massive neutrinos when extracting the CMB distance priors and find different values compared to those derived using the flat $\Lambda$CDM model. This clearly shows that the usual CMB distance priors are not sufficient to fully describe the physics when neutrino parameters are included. In Fig.\ \ref{fig:nuCDM_con_old}, we compare the constraints on the $\nu$CDM neutrino model (flat $\Lambda$CDM with two extra parameters $N_{\text{eff}}$ and $\sum m_{\nu}$) from the usual CMB distance priors and the full CMB data. In particular, we extract the CMB distance priors assuming a flat $\Lambda$CDM model and assuming the $\nu$CDM model and then put constraints on the $\nu$CDM model parameters using both sets of derived CMB distance priors. The results clearly show that the usual CMB priors fail to reproduce the constraints from the full CMB data, regardless of the model used in the extraction of the CMB distance priors. Note that assuming the correct model ($\nu$CDM) in the extraction of CMB distance priors does {\bf not} improve the constraints and the resulting constraints on the cosmological parameters are less constraining than those determined using the $\Lambda$CDM model distance priors (this is due to larger uncertainty in the $\nu$CDM model distance priors). There are several reasons for these results. First, adding the neutrino parameters can alter the determination of the size of the sound horizon from CMB data \cite{Elgaroy_2007}, and the resulting CMB shift parameters. Second, the number of free parameters in the $\nu$CDM model is larger than the dimension of the data vector, however this may not be a dominant factor since the various dynamical dark energy models and primordial power spectrum models used in the previous test had a similar number of degrees of freedom but the constraints were much less affected. In addition, we note that adding neutrino parameters can alter the degeneracy between the CMB shift parameters compared with the results derived using a flat $\Lambda$CDM model, which can also cause discrepancies in the final constraints.

\begin{figure}[htbp]
\begin{center}
\includegraphics[width=8.5cm]{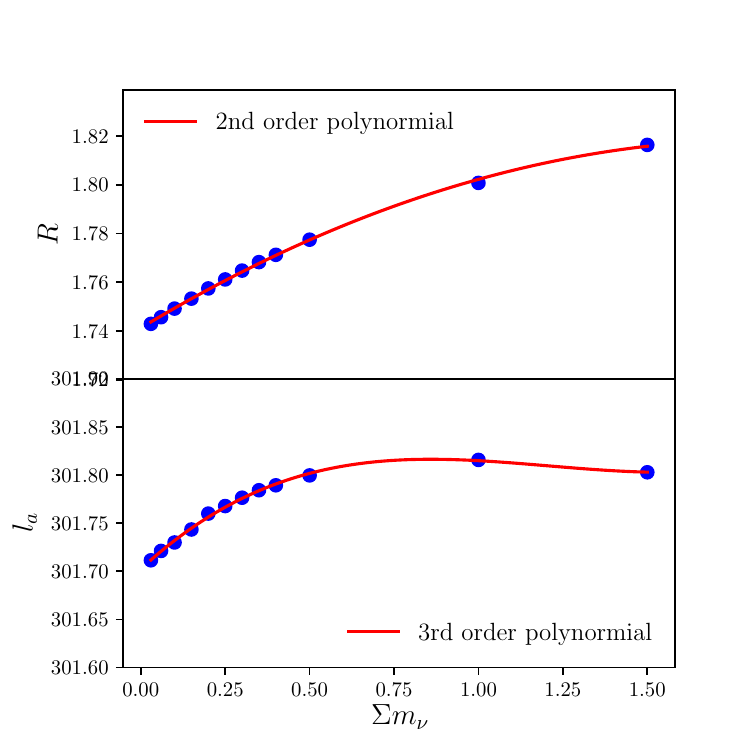}
\includegraphics[width=8.5cm]{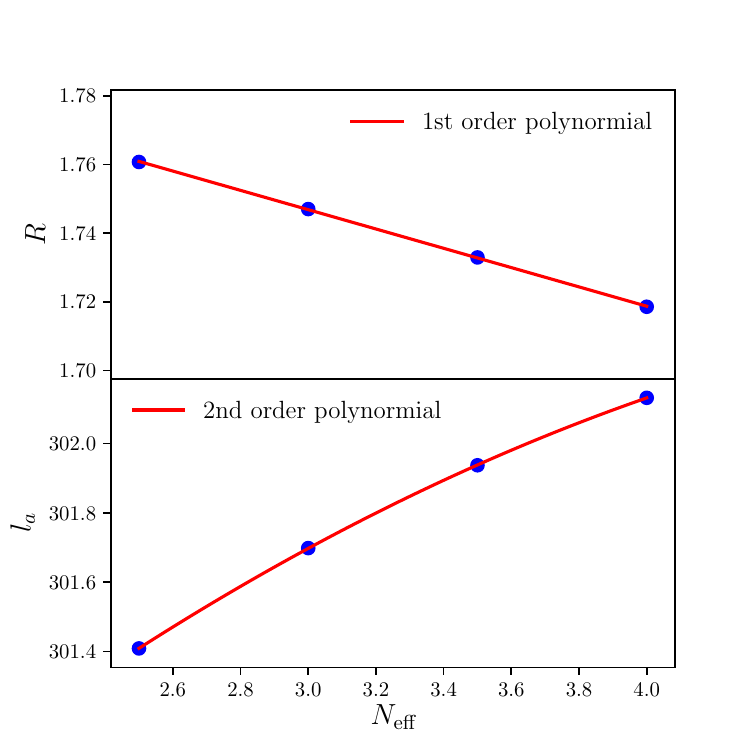}
\caption{One-dimensional dependence of the CMB shift parameters on the neutrino parameters, $top:$ neutrino mass $\sum m_{\nu}$, $bottom:$ effective number of relativistic species $N_{\text{eff}}$. For each data point, we fix all the other parameters at the flat $\Lambda$CDM model values. The red solid lines are simple polynomial fits to the results.}
\label{fig:nu_1Dfit}
\end{center}
\end{figure}

\begin{figure}[htbp]
\begin{center}
\includegraphics[width=8.5cm]{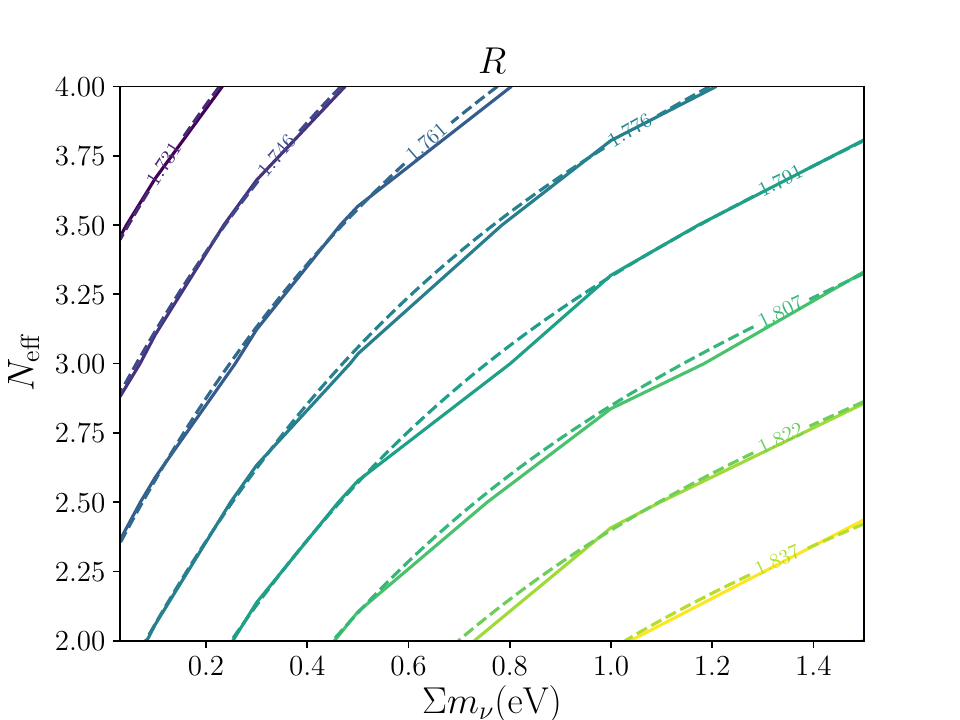}
\includegraphics[width=8.5cm]{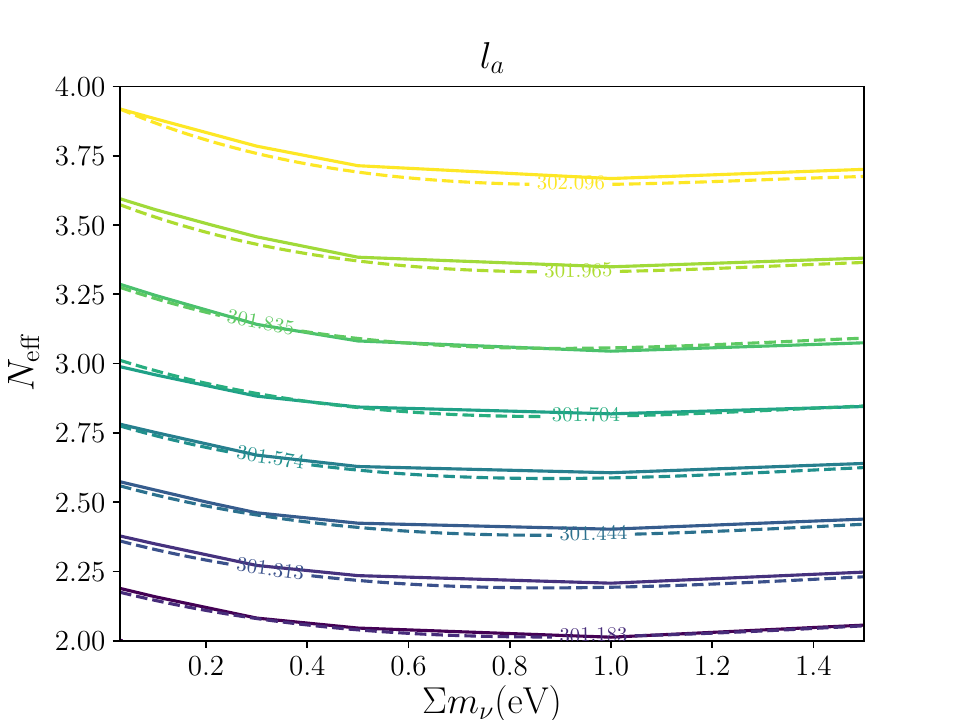}
\caption{Two-dimensional dependence of the CMB shift parameters on the neutrino parameters $\sum m_{\nu}$ and $N_{\text{eff}}$, $top:$ $R$; $bottom:$ $l_{a}$. The dashed lines are two-dimensional polynomial fits of third order, shown as contours. The solid lines are also contours, but measured directly from the data.}
\label{fig:nu_2Dfit}
\end{center}
\end{figure}

\begin{figure}[htbp]
\begin{center}
\includegraphics[width=8.5cm]{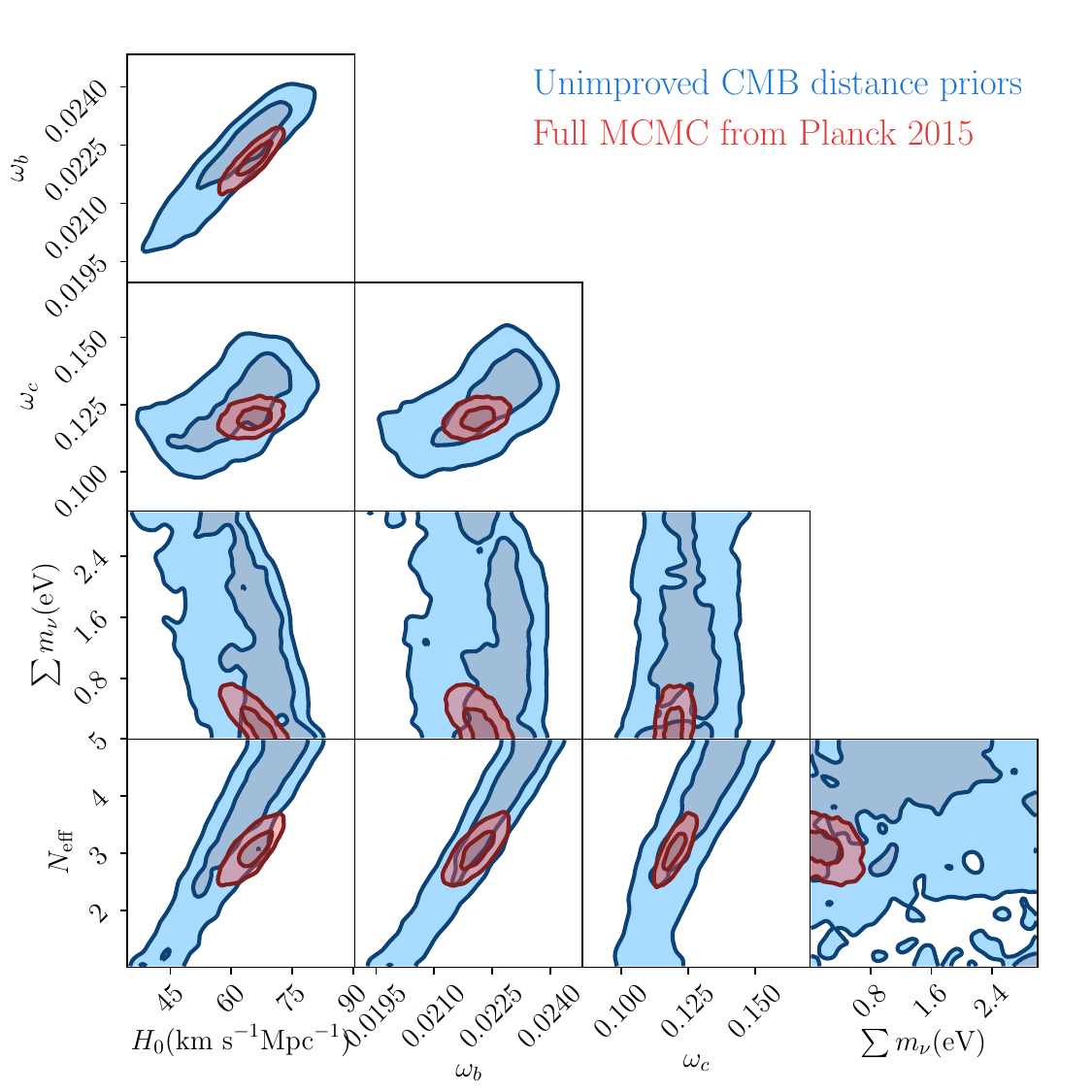}
\caption{Constraint on the $\nu$CDM model with the unimproved neutrino model-dependent CMB distance priors. For comparison, the result from the full Planck data is shown as the red contour.}
\label{fig:model_dependence_neutrino_constraint}
\end{center}
\end{figure}

The sum of neutrino masses and the number of relativistic species can be explicitly modeled in the CMB computation. This makes a model-dependent modeling of the CMB distance prior for the $\nu$CDM model possible. We investigate this possibility by running a grid of models with different $\sum m_{\nu}$ and $N_{\text{eff}}$ values. The results show a clear pattern for the CMB shift parameters in both the one-dimensional and two-dimensional cases, which can be accurately modelled by a simple polynomial fits of up to third order. We present these results in Figs.\ \ref{fig:nu_1Dfit} and \ref{fig:nu_2Dfit}. In these computations all the other parameters are fixed at the flat $\Lambda$CDM model values and only one or two neutrino parameters are changed each time. This simple dependence can easily improve the CMB distance priors dependence on $\sum m_{\nu}$ and $N_{\text{eff}}$. In addition, this also enables a neutrino model dependent modeling of the correlation between the CMB shift parameters, but the result is noisier than the shift parameters themselves. Using this model-dependent modeling of the CMB distance priors, we ran a MCMC analysis. However, it turns out that this method is also unable to provide constraints on the neutrino parameters that are consistent with those derived using the full CMB data, as shown in Figure \ref{fig:model_dependence_neutrino_constraint}. This implies that more prior information needs to be included for the neutrino model, not only this model dependence.

\subsection{Adding neutrinos}
\label{sec:neutrinos}

Since the sum of neutrino masses and $N_{\text{eff}}$ have multiple effects on the CMB, it is impossible to find a simple parameter to express the whole effect. However, for the CMB shift parameters, one of the most important factors is the correct modeling of the sound horizon; examples can be found in the fitting formula in Ref.\ \cite{Aubourg_2015}. Thus adding one or two neutrino-related parameters might be useful. In order to express the compressed CMB data in a Gaussian distributed form, to be consistent with the elegant and traditional expression, we extract additional information for the distribution of $\omega_{c}$ and $N_{\text{eff}}$ from the MCMC analysis. These boost the CMB distance prior set to be $(R, l_{a}, \omega_{b}, \omega_{c}, N_{\text{eff}})$. The resulting data vector and covariance matrix are

\begin{equation}\label{eq:vector}
    \mathbf{v}\equiv \begin{pmatrix}
    R\\
    l_a\\
    \omega_b\\
    \omega_c\\
    N_{\textsf{eff}}
    \end{pmatrix}=\begin{pmatrix}
    1.7646\\
    301.7790\\
    0.02212\\
    0.1201\\
    3.06756
    \end{pmatrix}
\end{equation}
\begin{eqnarray*}
    &&C_{\mathbf{v}}=10^{-8}\times \\
    &&\begin{pmatrix}
    45272.11~& -48046.40~& -691.72~& -294.67~& -364590.88 \\
    -48046.40~&  4117094.71~&  2729.81~&  61787.61~& 4529565.27 \\
    -691.72~&  2729.81~&  16.79~&  54.66~& 9753.96 \\
    -294.67~&  61787.61~&  54.66~& 1497.16~& 94816.70 \\
    -364590.88~&  4529565.27~&  9753.96~&  94816.70~& 9548751.16 \\
    \end{pmatrix}.
\end{eqnarray*}

\begin{figure}[htbp]
\begin{center}
\includegraphics[width=9.0cm]{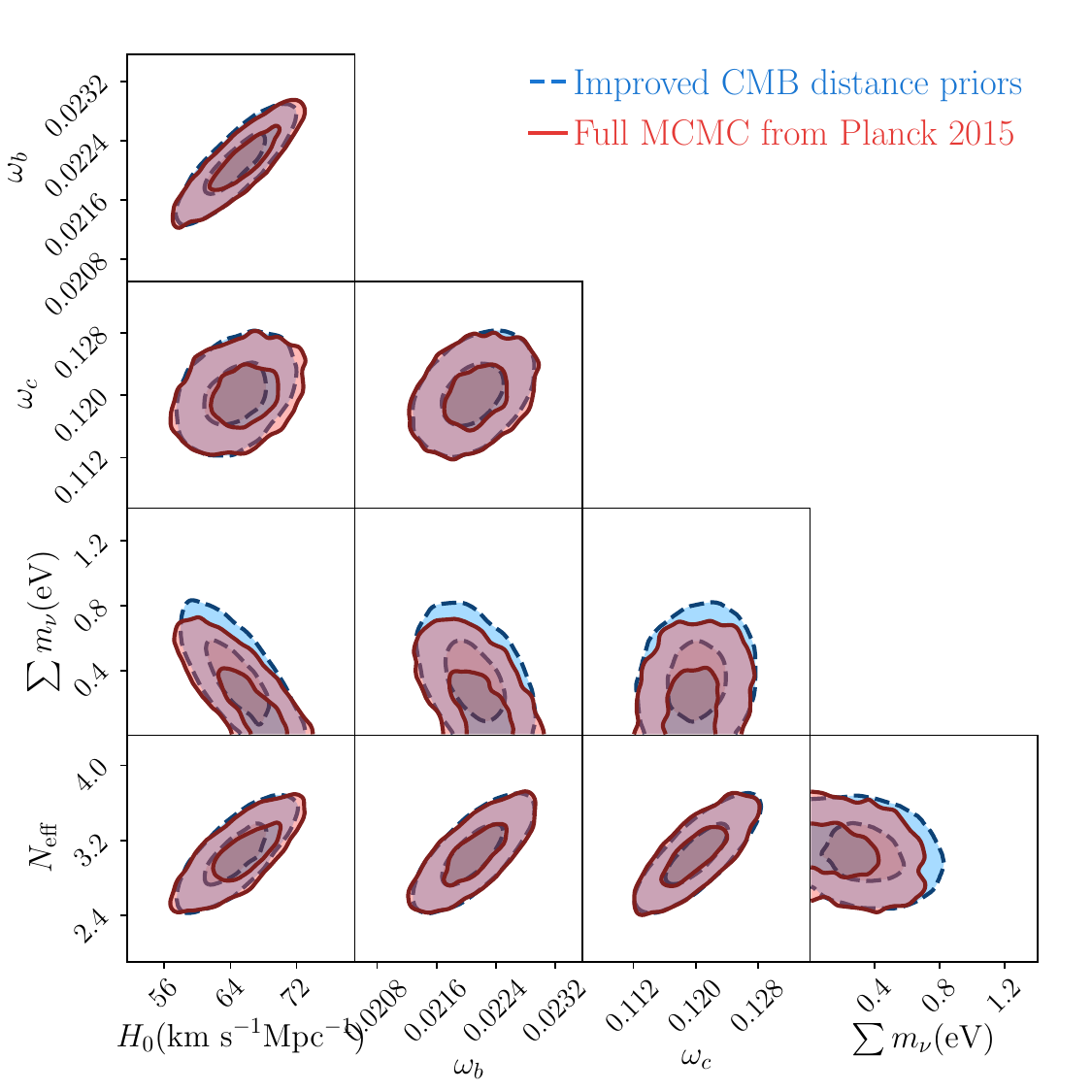}
\caption{Updated constraint on the $\nu$CDM model parameters. The CMB distance priors are extracted using the $\nu$CDM model to be self-consistent, and expressed in terms of five observables $(R, l_{a}, \omega_{b}, \omega_{c}, N_{\text{eff}})$ for the geometrical only aspect. The consistency with the full CMB data constraint validates our generalization of the CMB distance priors.}
\label{fig:nuCDM_con_new}
\end{center}
\end{figure}

Using the above in the usual recipe for including CMB distance priors in a likelihood analysis, e.g., \cite{Wang_2006, Wang_2007, Chen_2018, Zhai_2019, Arjona_2019}, we update the constraints on the $\nu$CDM model and compare with the full CMB data constraint in Fig.\ \ref{fig:nuCDM_con_new}. Clearly, this new set of CMB distance priors results in parameter constraints that are consistent with those derived using the full CMB data. The largest deviation is in the neutrino mass, but this is still smaller than $1\sigma$. This deviation is partly due to the effect of the correlation between the neutrino mass and other parameters, and partly due to the assumed Gaussian form of the data vector, both of which can slightly shift the upper end of the constraint to a higher mass. This result demonstrates that the CMB distance priors can be improved to correctly describe and constrain cosmological models with massive neutrinos by including constraints on the cold dark matter density $\omega_{c}$ and the effective number of relativistic species $N_{\text{eff}}$. 

We have used Planck 2015 data to arrive at our results so far, to make use of our extensive previous work. We have shown that the CMB distance priors need to be expanded to include parameters $\omega_c$ and $N_{\text{eff}}$ to be generally applicable in constraining models that allow the neutrino parameters to vary. 
With the final release of Planck observations \cite{Planck_2018_6}, we also extract the corresponding distance priors valid for neutrino models, which should be used in summarizing CMB data in a joint cosmological data analysis. The resulting data vector and covariance matrix are
\begin{equation}
    \mathbf{v}\equiv \begin{pmatrix}
    R\\
    l_a\\
    \omega_b\\
    \omega_c\\
    N_{\textsf{eff}}
    \end{pmatrix}=\begin{pmatrix}
    1.7661\\
    301.7293\\
    0.02191\\
    0.1194\\
    2.8979
    \end{pmatrix}
\end{equation}
\begin{eqnarray*}
    &&C_{\mathbf{v}}=10^{-8}\times \\
    &&\begin{pmatrix}
    33483.54~& -44417.15~& -515.03~& -360.42~& -274151.72 \\
    -44417.15~&  4245661.67~&  2319.46~&  63326.47~& 4287810.44 \\
    -515.03~&  2319.46~&  12.92~&  51.98~& 7273.04 \\
    -360.42~&  63326.47~&  51.98~& 1516.28~& 92013.95 \\
    -274151.72~&  4287810.44~&  7273.04~&  92013.95~& 7876074.60 \\
    \end{pmatrix}.
\end{eqnarray*}
We have used the chain \textmd{base\_nnu\_mnu\_plikHM\_TT\_lowl\_lowE\_post\_lensing} from the Planck final data results in the Planck archive for deriving the above CMB distance priors. 

\begin{figure}[htbp]
\begin{center}
\includegraphics[width=8.5cm]{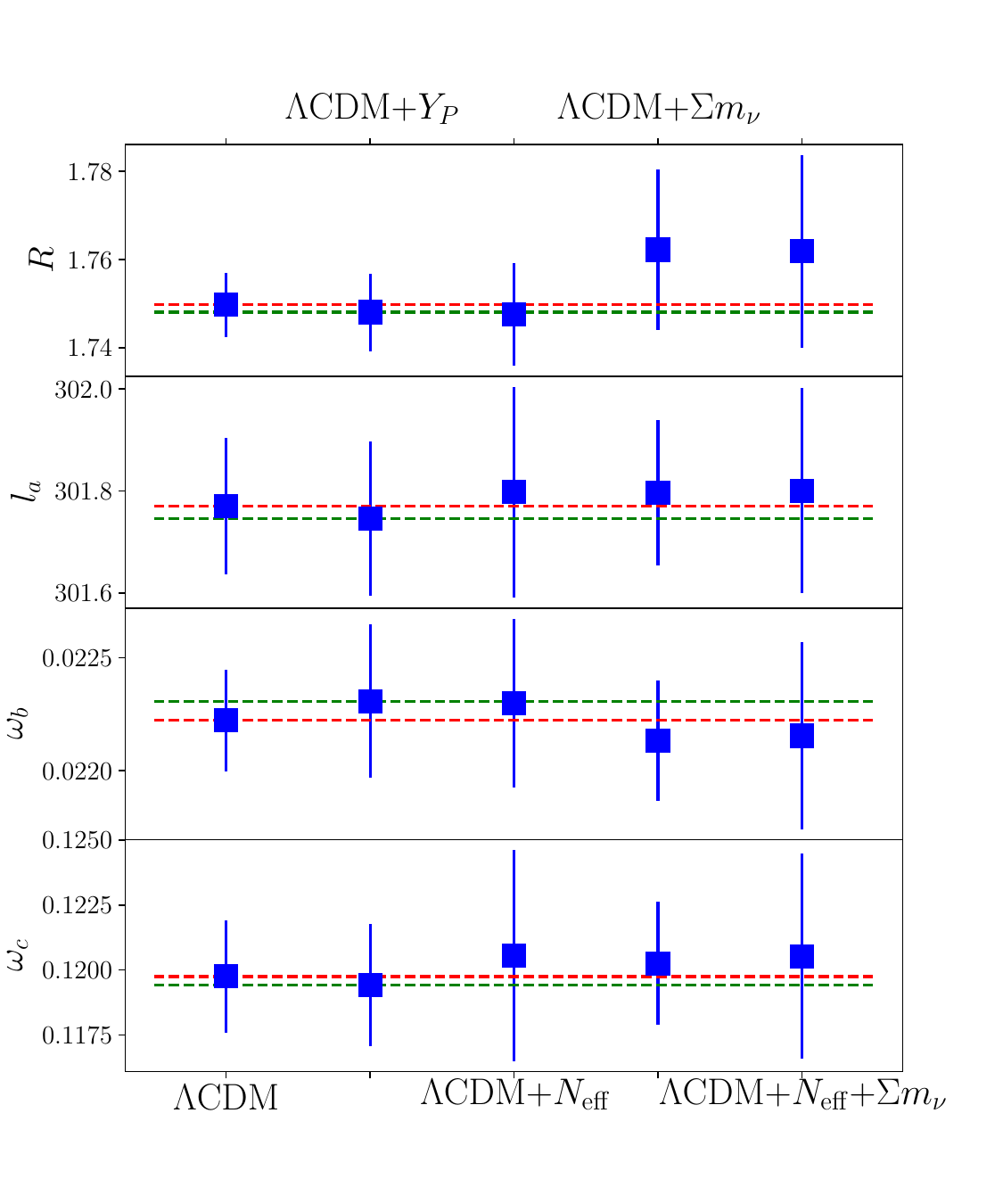}
\caption{The CMB distance priors $R, l_{a}$, $\omega_{b}$ and $\omega_{c}$ extracted by assuming different models as indicated on the $x$-axes, but from the same input CMB \textsf{TT+lowP} data. The red and green dashed lines indicate the spatially flat $\Lambda$CDM model with tilted primordial power spectrum and the tilted flat $\Lambda$CDM model with varying $Y_p$, respectively.}
\label{fig:CMBYhe}
\end{center}
\end{figure}

Since the effect on the CMB distance priors from the neutrino parameters alters the evolution of the early universe, some other physical mechanisms may have similar impact due to the degeneracy with the neutrino parameters. We investigate this by testing another extension of the $\Lambda$CDM model which allows the helium fraction $Y_{P}$ to vary as a free parameter. The distance priors of this model is extracted from the already existed MCMC chain from Planck 2015 release. In particular, we compare the result with the standard $\Lambda$CDM model and $\nu$CDM model from \textsf{TT+lowP} data. We present the resulting CMB distance priors in Figure \ref{fig:CMBYhe}. We can see that the measurement of the CMB distance priors are similar and roughly consistent within $1\sigma$ compared with $\Lambda$CDM model. However, the uncertainties are affected significantly, as well as the covariance between observables. Therefore it is possible that the traditional form of the CMB distance priors $(R, l_{a}, \omega_{b})$ may not be able to provide unbiased constraints on cosmological models with varying helium fraction. One possible solution might be similar to what we used for the neutrino model, incorporating the measurement of $Y_{P}$ as additional prior information.

\section{Discussion and conclusion}\label{sec:conclusion}

As one of the least model-dependent quantities that can be extracted from the CMB power spectrum, the CMB distance priors can represent nearly all of the CMB information relevant for probing dark energy. 
They can also lead to a better understanding of the CMB constraints on model parameters, including the uncertainties and correlations with other cosmological measurements, without computing the full linear perturbation theory CMB quantities. Their usage has been demonstrated in many applications and it has provided valuable results. However, the CMB distance priors are not completely model-independent, from both their definition and the way they are measured. Thus it is possible that some information of the CMB data is lost in this data compression process. The application to arbitrary models, especially models with radical differences compared with the standard flat $\Lambda$CDM model, may lead to non-negligible bias in the estimated parameter values. 

In this work we revisit the usage of the CMB distance priors to constrain various cosmological models. With the same input CMB data from Planck, we first extract and compare the CMB distance priors assuming different cosmological parameters, including those that  describe dark energy dynamics, spatial curvature, the primordial power spectrum, and neutrino properties. Our results show that for many models that alter the dynamics of dark energy, or generalize the simple tilted primordial power spectrum model, the CMB distance priors values are not significantly affected. This implies that the CMB distance priors are a faithful substitute for the full CMB data and can provide consistent and unbiased constraints for these models. The biggest deviation we observe is for the untilted primordial power spectrum due to non-zero spatial curvature. This untilted non-flat model is not nested with the simple tilted flat model. We also note that this model has a significantly larger minimum $\chi^{2}$ value which indicates a systematic offset of all the cosmological parameters. This leads to a change in the CMB distance priors and care must be taken when using the CMB distance priors to analyze such models.

We further examine the CMB distance priors when neutrino parameters are varied. Although these models are nested with the standard $\Lambda$CDM model, the usual CMB distance priors are not sufficient to provide constraints consistent with those from the full CMB data. Since the effect of the sum of neutrino masses and additional relativistic species can be computed explicitly for the CMB, we explore a possible solution by implementing a model-dependent CMB distance prior. However, it turns out that this method is not able to fix the problem. This implies that a direct modeling of the CMB physics in the CMB distance priors must be considered. We thus add the constraints on the parameters $\omega_{c}$ and $N_{\text{eff}}$ as new data points in the set of CMB distance priors. With this expansion, we find that the constraints on the $\nu$CDM model from the CMB distance priors are consistent with the results from the full CMB data. The only deviation is in the constraint on the neutrino mass, which is still smaller than 1$\sigma$. Therefore this result validates our new modeling of the CMB distance priors, which now can be applied to cosmological models with free neutrino parameters.

The CMB distance priors are a simple and physical method of compressing the CMB data. However this method is not exclusive and other methods are also worth investigating, for instance the method discussed in Ref.\ \cite{Prince_2019}. CMB data compression is not only useful for dimension reduction of the data vector and covariance matrix, but also helps in finding model-independent observables relevant for CMB observations. This can provide quick and accurate tests for non-standard cosmologies. 
In addition, although there is an apparent tension of the $H_{0}$ measurement from the CMB and from some local distance ladder data, it is worthwhile to note that the value from the CMB is dependent on the underlying model assumed. Therefore the uncertainty in this measurement may be underestimated.
A model independent method to analyze the CMB data is important in this regard and might provide hints of a possible underlying physics explanation.

\acknowledgments

We thank the anonymous referee for the valuable comments and suggestions that have helped us improve the contents of this paper. We acknowledge the use of the public softwares CAMB \citep{Lewis_2002ah}, Matplotlib \citep{matplotlib}, NumPy \citep{numpy}, SciPy \citep{scipy}, ChainConsumer \citep{Hinton_2016}, Scikit-learn \citep{scikit-learn} and Emcee \citep{Foreman-Mackey_2013}. This work is supported in part by NASA grant 15-WFIRST15-0008, Cosmology with the High Latitude Survey WFIRST Science Investigation Team (SIT). C.-G.P.\ was supported by the Basic Science Research Program through the National Research Foundation of Korea (NRF) funded by the Ministry of Education (No. 2017R1D1A1B03028384). B.R.\ was supported in part by DOE grant DE-SC0019038

\bibliography{DE_GoF,software}

\end{document}